\documentclass[showpacs,prb,preprint]{revtex4}
\usepackage{amssymb}
\usepackage{graphicx}

\begin{document}

\title{Graphene spin capacitor for magnetic field sensing}
\author{Y. G. Semenov, J. M. Zavada, and K. W. Kim}
\address{Department of Electrical and Computer Engineering, North Carolina
State University, Raleigh, NC 27695-7911}

\begin{abstract}
An analysis of a novel magnetic field sensor based on a graphene spin
capacitor is presented. The proposed device consists of graphene nanoribbons
on top of an insulator material connected to a ferromagnetic source/drain.   The time evolution of spin polarized electrons injected into the capacitor
can be used for an accurate determination at room temperature of external
magnetic fields. Assuming a spin relaxation time of 100~ns, magnetic
fields on the order of $\sim 10$ mOe may be detected at room temperature. The observational accuracy of this device depends on the density of magnetic defects and spin
relaxation time that can be achieved.
\end{abstract}

\pacs{85.75.-d,72.25.-b,75.70.-i,73.61.Wp}

\maketitle


During the past two decades research in spintronic applications has focused
on imitating conventional electronics based on charge transport and
manipulation. A typical example is the Datta-Das spin field effect
transistor,~\cite{Datta90} which generated a wide range of variations~\cite%
{Zutic04} albeit without practical success. In general, spin transistors
require (i) a long spin relaxation time $\tau _{s}$ compared to the spin
processing time in the semiconductor channel, (ii) a sufficient spin
polarization that allows discrimination between the "on" and "off" states,
and (iii) stability against electronic thermal dispersion. Note that spin
manipulation via the Bychkov-Rashba effect~\cite{Bychkov84} relies on the
strength of spin-orbit coupling, which in turn is incompatible with long
spin relaxation.

This Letter considers the possibility of separating the spin and charge
processing in a capacitor populated with spin-polarized electrons. Robust
functionality of such a spin capacitor relies on tolerance to spin injection
efficiency and subsequent phase (rather than spin polarization magnitude)
manipulation of spin polarization. Here we demonstrate that the state-of-art
achievements in injection, detection and storage of the electron spins in a
graphene based capacitor can lead to a class of spintronic devices designed
for detection at room temperature of extremely weak magnetic fields.

A schematic of the proposed device is shown in Fig.~1. The device consists
of graphene nanoribbons situated on top of an insulator material connected
to a ferromagnetic (FM) source/drain. This structure forms a capacitor,
which stores the electronic charge $Q$ and spin under the appropriate bias $%
V_{g}$ conditions. Operation of the spin capacitor sensor is depicted in
Fig.~2. After electron injection through the FM contact a spin polarization $%
\mathbf{P(}t\mathbf{)}$ is established at $t=0$ [see Fig.~2(a)]. Unlike
charge storage, $\mathbf{P(}t\mathbf{)}$ changes in time due to spin
precession in an external magnetic field and spin decoherence. As the
electric bias is reversed at $t=\tau _{ex}$, the electrons leave the capacitor
with rotated and reduced spin polarization so that the peak intensity $%
I_{peak}$ of probe current $I_{prob}$ through the FM contact depends on the
magnitude of $P\mathbf{(}t\mathbf{)}$ and phase of its rotation [Fig.~2(b)].
The temporal oscillations of $\mathbf{P(}t\mathbf{)|}_{t=\tau _{ex}}$ can be
recorded by applying a series of sequential measurements with variable
exposure times $\tau _{ex}$, i.e., the duration of the electron storage in
the capacitor [Fig.~2(c)]. Moreover, the iteration of such a procedure in
three orthogonal device directions will supply complete information about
the strength and direction of an external magnetic field.

The advantage of the proposed device consists in an easy electrical
variation of spin exposure time and elimination of spin dephasing that
arises in transient two-terminal schemes involving transport. The conducting
material for such a spin capacitor must possess a high electron mobility and
long spin relaxation time including semiconductors with weak spin-orbital
interaction. The best candidate for this application appears to be graphene
nanoribbons with tunable band-gap, low spin-orbital coupling and high
electron mobility.~\cite{Neto09}

In general, $\mathbf{P(}t\mathbf{)}$ is the difference between the
ensemble-averaged spin polarization at time $t$ [i.e., $\left\langle \mathbf{%
S}\right\rangle (t)$] and that of the equilibrium value $\left\langle
\mathbf{S}\right\rangle _{0}$. Its time evolution can be described by the
Bloch equation
\begin{equation}
\frac{d\mathbf{P(}t\mathbf{)}}{dt}=\mathbf{\varpi }\times \mathbf{P(}t%
\mathbf{)-}\frac{\mathbf{P(}t\mathbf{)}}{\tau _{s}}  \label{1}
\end{equation}%
with initial condition $\mathbf{P(}0\mathbf{)=P}_{0}$, where $\mathbf{P}%
_{0}\parallel \mathbf{M}$ originates from spin injection from a FM contact
with magnetization $\mathbf{M}$ and $\tau _{s}$ is the characteristic spin
relaxation time mentioned earlier. To increase the accuracy of the device,
the internal magnetic field $\mathbf{H}_{in}$ due to the magnetic parts of
the device must be incorporated into consideration along with the external
contribution $\mathbf{H}_{ex}$.~\cite{comm1} Thus, the electron spin
precession vector is given by $\mathbf{\varpi =}\gamma _{e}\mathbf{H}$,
where $\mathbf{H}=\mathbf{H}_{in}+\mathbf{H}_{ex}$ and $\gamma _{e}$ is is
the electron gyromagnetic ratio. For simplicity, we assume that magnetic
fields are not too strong to require distinguishing between longitudinal and
transversal spin relaxation times. Then the solution of Eq.~(\ref{1}) reads
\begin{equation}
\mathbf{P(}t\mathbf{)}=P_{0}e^{-t/\tau _{s}}\left\{ \mathbf{p}\cos \omega t+2%
\mathbf{n}(\mathbf{n\cdot p})\sin ^{2}\frac{\omega t}{2}+ (\mathbf{n}\times \mathbf{p}) \sin \omega t\right\} \,,  \label{2}
\end{equation}%
where $P_{0}=\left\vert \mathbf{P}_{0}\right\vert $, $\mathbf{p=P}_{0}/P_{0}$, $\omega =\left\vert \mathbf{\varpi }\right\vert $, and $\mathbf{n=\varpi /}%
\omega =\mathbf{H}/H$. The spin-dependent output signal from the FM contact is given as $I_{peak}(t)\propto \mathbf{P}(t)\cdot \mathbf{p}$ in
terms of variable exposure time $t=\tau _{ex}$.

Once $I_{peak}(t)$ is measured in the manner discussed above, the Fourier
transformation $F(f)=\int\nolimits_{0}^{\infty }I_{peak}(t)\cos
(f t)dt$ can be found as
\begin{equation}
F(f )=\tau _{s}P_{0}^{2}\left( \frac{\cos ^{2}\alpha }{1+x^{2}}+\frac{%
(1+x^{2}+y^{2})\sin ^{2}\alpha }{[1+(x+y)^{2}][1+(x-y)^{2}]}\right) ,
\label{3}
\end{equation}%
where $x=\tau _{s}f $, $y=\tau _{s}\omega $, and $\alpha $ is the angle
between $\mathbf{p}$ and $\mathbf{n}$. In the case of $\alpha =\pi /2$
(i.e., the magnetic field is perpendicular to the injected spin polarization $
\mathbf{P}_{0}$) at $f =0$ (the dc component), Eq.~(\ref{3})
reproduces the Hanle effect: $F(0)=\tau _{s}P_{0}^{2}/(1+\omega
^{2}\tau _{s}^{2})$,~\cite{DP84} which can be applied to the determination of
prefactor of Eq.~(\ref{3}) and relaxation time $\tau _{s}$.  More importantly, $F(f)$ possesses a pronounced peak at $\omega $ provided $ 1/\tau _{s}\ll \omega $ and the angle $\alpha $ is not close to zero.
Then the peak position $f_p$ immediately indicates the strength of the total magnetic field $H= f_{p}/\gamma _{e}$. Note that this treatment does not
involve determination of the peak intensity making the sensor insensitive to
spin injection efficiency.

In the case of non-negligible $1/\tau _{s} \omega$ or slight
deviation $\mathbf{H}$ from the direction of spin polarization $\mathbf{P}%
_{0}$ (small $\alpha$), the peak of $F(f )$ can be masked and shifted from the resonant frequency $\omega $ by the first term in Eq.~(\ref{3}). To mitigate these effects, the signal processing can be modified by subtracting $2\tau
_{s}P_{0}^{2}/(1+x^{2})$ from $F(f )$.  This procedure defines a new
function $\Phi (f )=2\tau _{s}P_{0}^{2} \phi _{S}(x,y) \sin ^{2}\alpha
$, where $\phi _{S}(x,y)$ takes the form of the expression in brackets of
Eq.~(\ref{3}) after substitutions $\cos ^{2}\alpha \rightarrow -1$ and $\sin
^{2}\alpha \rightarrow 1$.  Even though the peak position $ f_{p}$ of $\Phi (f ) [\sim \phi _{S}(x,y)$] is now no longer dependent on $\alpha$, it is still subject to deviation from the desired answer (i.e., $\omega$) as $1/\tau _{s} \approx \omega$ in Fig.~3.  Analysis
indicates that this error can be corrected by equation%
\begin{equation}
\omega \tau_s = y=x_{p}-\frac{b}{6x_{p}+x_{p}^{3}},  \label{3a}
\end{equation}%
which holds with high precision for $f_p\tau_s \equiv x_{p}\gtrsim 1$ and $b=2.5$. Thus,
Eqs.~(\ref{3}) and (\ref{3a}) coupled with function $\Phi (f )$
establish the basis for the spin capacitor application to magnetic field
measurement.

In order to determine the external field $\mathbf{H}_{ex}$, measurement of $
\mathbf{H}$ (or, more practically, $\omega$ via $f_p$) must be completed in at least three different orientations of the device, e.g., along the X, Y and Z axes. Apparently, repositioning the device does not alter the contribution
of $\mathbf{H}_{in}=(H_{\mathrm{X}}^{(in)},H_{\mathrm{Y}}^{(in)},H_{\mathrm{Z%
}}^{(in)})$ but results in permutation of $\mathbf{H}_{ex}$ components as
\begin{eqnarray}
&&\mathbf{H}^{(1)}=\left( H_{\mathrm{X}}^{(ex)},H_{\mathrm{Y}}^{(ex)},H_{%
\mathrm{Z}}^{(ex)}\right) ,  \nonumber \\
&&\mathbf{H}^{(2)}=\left( -H_{\mathrm{Y}}^{(ex)},H_{\mathrm{X}}^{(ex)},H_{%
\mathrm{Z}}^{(ex)}\right) , \\
&&\mathbf{H}^{(3)}=\left( H_{\mathrm{X}}^{(ex)},-H_{\mathrm{Z}}^{(ex)},H_{%
\mathrm{Y}}^{(ex)}\right) .  \nonumber
\end{eqnarray}%
Solutions $y_{k}$ of Eq.~(\ref{3a}) for resonant frequencies $\omega _{k}=y_{k}/\tau _{s}$ that correspond to the different
device orientations $k=1,2,3$ (i.e., three equations) yield all three components of $\mathbf{H}_{ex}$ (three unknowns) assuming that $H_{j}^{(in)}$ are known parameters of the device:
\begin{equation}
\gamma _{e}^{2}\sum\limits_{j=\mathrm{X,Y,Z}}\left(
H_{j}^{(in)}+H_{j}^{(k)}\right) ^{2}=\omega _{k}^{2}\,.  \label{4}
\end{equation}

An important characteristic of the device is its detection sensitivity,
which is closely related to the accuracy $\varepsilon $ of measurement of the peak position $f_{p}$ for $\Phi (f)$.  Limiting the measurement error $\Delta f_p$ to a fraction of the intrinsic broadening ($1/\tau_s$) of the Fourier transformed signal, $\Delta f_p \tau_s = \gamma _{e}\Delta H \tau_{s} \approx\varepsilon \ll 1$. For example, $\varepsilon \sim$~1~\% leads to $\Delta H\sim 5\times 10^{-10}($Oe$)/\tau_{s}($s$)$. This estimate shows that the relaxation time $\tau _{s}$ is a crucial parameter, which limits the applicability of the device. Consequently, the spin relaxation time $\tau _{s}$ of electrons needs to be evaluated.

The most important mechanism at room temperature appears to be electron decoherence due to irregular interactions with localized spin moments as the electron thermally fluctuates between localized states.~\cite{SK04} For
electron localization in graphene nanoribbons in particular, magnetic
defects such as the ribbon edges or carbon vacancies could be responsible
for spin relaxation. Taking into account uncertainties of the interaction
strength and density of defects, we can evaluate the relaxation time $\tau
_{s}$ in a quantitative manner.

It has been shown that the fluctuation in local field $\hbar \mathbf{\Omega }%
(t)$ with characteristic time $\tau _{p}$ can be responsible for electron
spin relaxation rate $\tau _{s}^{-1}=\frac{1}{2}\left\langle \mathbf{\Omega }%
^{2}(t)\right\rangle \tau _{p}$.~\cite{SK04,Shklovskii06} We estimate $%
\left\langle \mathbf{\Omega }^{2}(t)\right\rangle $ assuming that phonon
mediated transitions within energy spectrum $\varepsilon _{\nu }$ of
nanoribbon confinement potential produce fluctuations of electron densities $%
\left\vert \Psi _{\nu }(r_{i})\right\vert ^{2}$ at the $N$ magnetic moments
located at lattice sites $r_{i}=(x_{i},z_{i})$, $i=1,...,N$. We also assume that the corresponding electron wave functions $\Psi _{\nu }(x,z)=\psi _{\nu
}(z)\phi _{0}(x)$ can be factorized into longitudinal $\psi _{\nu }(z)$ and
transversal $\phi _{0}(x)$ parts, where $\phi _{0}(x)$ is bounded by ribbon
width $d$. For such electron states, the effective field induced by magnetic
defects with spins $\mathbf{I}_{i}$ takes the form in energy units%
\begin{equation}
\hbar \mathbf{\Omega }_{\nu }=Ja_{0}\sum\limits_{i}\left\vert \psi _{\nu
}(z_{i})\right\vert ^{2}\left\vert \phi _{0}(x_{i})\right\vert ^{2}\mathbf{I}%
_{i},  \label{5}
\end{equation}%
where $J$ is the exchange interaction constant of the magnetic defect and $%
a_{0}$ is the area of graphene unit cell. At room temperature, a weak
magnetic field cannot polarize the localized spins so that a thermal
averaging yields $\left\langle \mathbf{I}_{i}\right\rangle =0$ and $%
\left\langle \mathbf{\Omega }_{\nu }\right\rangle =0$. On the other hand,
the random dispersion of non-correlated $\mathbf{I}_{i}$ leads to the finite
mean square result
\begin{equation}
\left\langle \mathbf{\Omega }_{\nu }^{2}\right\rangle =\frac{J^{2}a^{2}}{%
\hbar ^{2}}\sum\limits_{i}\left\vert \psi _{\nu }(z_{i})\right\vert
^{4}\left\vert \phi _{0}(x_{i})\right\vert ^{4}I(I+1).  \label{6}
\end{equation}

Electron thermal fluctuations among the different states $\nu $ leads to
dispersion of the spin precession $\left\langle \mathbf{\Omega }%
^{2}\right\rangle =\sum_{\nu }\left\langle \mathbf{\Omega }_{\nu
}^{2}\right\rangle P_{\nu }$, where $P_{\nu }=e^{-\varepsilon _{\nu
}/k_{B}T}/\sum_{\nu }e^{-\varepsilon _{\nu }/k_{B}T}$. Applying the
eigenfunctions $\psi _{\nu }(z)$ for the confined potential and summing over
$N$ localized spins by integrating with a linear density $n_{1}$, one can
find a mean-square estimate. In the simplest case of a strong confinement
for a range $L\gg d$ along nanoribbon imposes $\varepsilon _{\nu
}-\varepsilon _{d}=\pi ^{2}\hbar ^{2}\nu ^{2}/2m^{\ast }L^{2}$ with $\nu
=1,2,...$ and $m^{\ast }=\pi \hbar /v_{F}d$. Here, $\varepsilon _{d}$ is the
energy of transversal confinement. Under the condition that $\varepsilon
_{1}\ll k_{B}T\ll \varepsilon _{d}$, averaging of Eq.(\ref{6}) over $\nu $
can be approximated with
\begin{equation}
\left\langle \mathbf{\Omega }^{2}\right\rangle =\frac{9}{4\pi }\frac{%
J^{2}a^{2}n_{1}}{\hbar ^{2}d^{2}L}I(I+1)\ln \frac{2mL^{2}k_{B}T}{\pi
^{2}\hbar ^{2}}.  \label{7}
\end{equation}

The two-level model in Ref.~(\onlinecite{SK04}) predicts slowing of spin
relaxation in the limit of high temperatures due to a dynamical averaging
effect. However, the rate of spin relaxation in a multilevel system slightly
increases with temperature. Using typical parameters for graphene $J=0.1$
eV, $I=1/2$, $\tau _{p}=1$ ps and for nanoribbons with dimensions, $d=20$ nm
and $L=1$~$\mu $m, Eq.~(\ref{7}) leads to a spin relaxation time estimate of
$\tau _{s}\simeq $100 ns provided the density of non-compensated spins is $%
n_{1}\simeq 10^{6}$ cm$^{-1}$. For this case, the graphene spin capacitor can
determine magnetic fields as low as 10 mOe at room temperature assuming measurement uncertainty $\varepsilon$ of a few percent.

In the limiting case of zero magnetic impurity concentration, the weak
hyperfine interaction with $^{13}$C (interaction constant of approximately $-
$44 MHz)~\cite{Yazyev08} leads long spin relaxation times $\tau _{s}\gg 1$%
~s. Another mechanism due to surface irregularities, adapted to electron
diffusion over a discrete energy spectrum,~\cite{Brataas07} predicts more
realistic values of $\tau _{s}\sim $ 100 $\mu $s. Thus purification of
graphene nanoribbons can enhance the room temperature sensitivity of the
spin capacitor device to a few $\mu $Oe.

In summary, a concept for a magnetic sensing spintronic device is proposed,
based on spin phase measurements of a graphene spin capacitor that provides
greater sensitivity at room temperature than conventional magnetic sensors.~%
\cite{Ledbetter08} In addition, this device is expected to possess excellent
scalability and can be easily integrated to the current electronics
technology.

This work was supported in part by the US Army Research Office, NSF, and the
FCRP Center on Functional Engineered Nano Architectonics (FENA).

\newpage \noindent Figure captions

\vspace{0.2in} \noindent Fig. 1. (Color Online) Schematic view of the spin
capacitor magnetic sensor which consists of graphene nanoribbons on top of
an insulator material connected to a ferromagnetic source/drain with
magnetization vector $\mathbf{M}$. Under an applied bias, polarized electrons
are injection into the capacitor and their spins precess around the vector
sum $\mathbf{H}$ of external $\mathbf{H}_{ex}$ and internal $\mathbf{H}_{in}$
magnetic fields. The gate voltage pulses $V_{g}$ control electron expose
time $\tau _{ex}$. The current response $I_{prob}$ is used to determine $%
\mathbf{H}_{ex}$.

\vspace{0.2in} \noindent Fig. 2. Diagram of (a) the applied voltage pulses $%
V_{g}(t)$ and (b) current response $I_{prob}(t)$ upon a reverse pulse. The
peak amplitude $I_{peak}$ is phase-specific such that it is maximal for spin
polarization parallel to $\mathbf{M}$, and minimal for opposite spin phase.
(c) Oscillations of $I_{peak}$ that are recorded after a series of
measurements at different $\tau _{ex}$. The frequency determines the
strength of the total magnetic field. The spin relaxation time leads to
damping of the oscillations.

\vspace{0.2in} \noindent Fig. 3. Function $\phi _{S}(x,y)$ plotted at
different $y$ (=$\tau_s\omega$). The dashed vertical lines indicate the maximum of each curve (i.e., $x_p = f_p\tau_s$).  As $1/y$ becomes non-negligible, $f_p$ deviates from the resonant frequencies $y/\tau _{s}=\gamma _{e}H$ requiring the use of Eq.~(\ref{3a}).

\newpage

\begin{center}
\begin{figure}[tbp]
\includegraphics[scale=1.,angle=0]{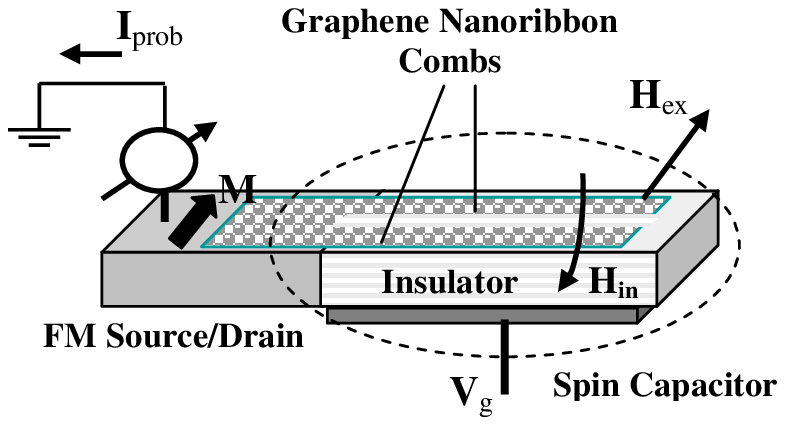}
\end{figure}
\vspace{320pt} {\large Fig. 1: Semenov et al. }
\end{center}

\newpage

\begin{center}
\begin{figure}[tbp]
\includegraphics[scale=1.,angle=0]{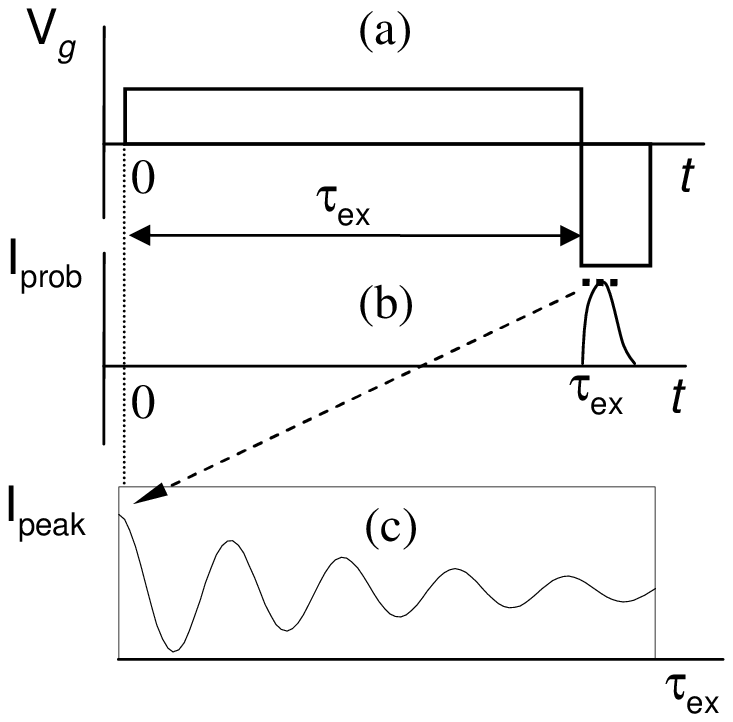}
\end{figure}
\vspace{320pt} {\large Fig. 2: Semenov et al. }
\end{center}

\newpage

\begin{center}
\begin{figure}[tbp]
\includegraphics[scale=1.,angle=0]{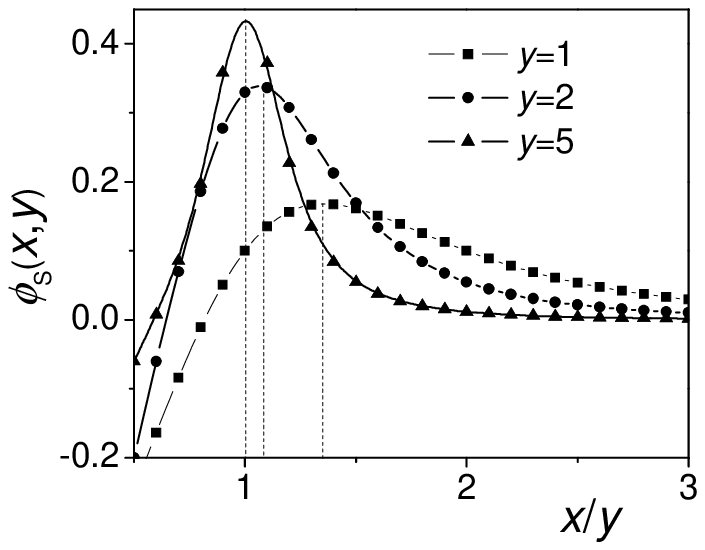}
\end{figure}
\vspace{320pt} {\large Fig. 3: Semenov et al. }
\end{center}


\begin{thebibliography}{99}
\bibitem{comm1} The effect of $\mathbf{H}_{in}$ on spin precession in the
channel (Fig.~1) may be negated, at least in part, by applying a subsidiary magnet with magnetization compensating that in the
FM source/drain.
\end{thebibliography}
\end{document}